# Coherent Order Parameter Oscillations in the Ground State of the Excitonic Insulator Ta$_2$NiSe$_5$

Daniel Werdehausen[1,2], Tomohiro Takayama[1,3], Marc Höppner[1], Gelon Albrecht[1,2], Andreas W. Rost[1,3], Yangfan Lu[4], Dirk Manske[1], Hidenori Takagi[1,3,4], and Stefan Kaiser[1,2,*]

[1]*Max-Planck-Institute for Solid State Research, 70569 Stuttgart, Germany*

[2]*4th Physics Institute, University of Stuttgart, 70569 Stuttgart, Germany*

[3]*Institute for Functional Matter and Quantum Technologies, University of Stuttgart, 70569 Stuttgart, Germany*

[4] *Department of Physics, The University of Tokyo, Bunkyo-ku, Tokyo 113-0033, Japan*

*Corresponding author. Email: s.kaiser@fkf.mpg.de

**The excitonic insulator is an intriguing electronic phase of quasi-condensed excitons. A prominent candidate is the small bandgap semiconductor Ta$_2$NiSe$_5$, in which excitons are believed to undergo a BEC-like transition. But experimental evidence for the existence of a coherent condensate in this material is still missing. A direct fingerprint of such a state would be the observation of its collective modes, which are equivalent to the Higgs- and Goldstone-modes in superconductors. Here we report evidence for the existence of a coherent amplitude response in the excitonic insulator phase of Ta$_2$NiSe$_5$. Using non-linear excitations with short laser pulses we identify a phonon-coupled state of the condensate that can be understood as a coupling of its electronic Higgs-mode to a low frequency phonon. The Higgs-mode contribution substantiates the picture of an electronically driven phase transition and characterizes the transient order parameter of the excitonic insulator as a function of temperature and excitation density.**

Excitonic insulators are expected in semimetals with a small band-overlap or small bandgap semiconductors [1-5]. In semiconductors a Bose-Einstein-like condensation of preformed excitons can occur if their binding energy exceeds the bandgap, whereas semimetals undergo a BCS-like transition [6]. A characteristic spectroscopic fingerprint of this transition is a band flattening, which can be observed in ARPES measurements [7-9]. By itself, however, this cannot elucidate the microscopic origin of the gap of a potential excitonic insulator. For this purpose ultrafast methods provide the possibility to gain further insight. Gap-melting times observed in time-resolved ARPES measurements were used to distinguish Mott, excitonic, and lattice driven dynamics [10,11]. In particular, for the charge density wave and the excitonic order in $TiSe_2$ a combination of time resolved ARPES [11, 12], x-ray [13], electron diffraction [14], as well as THz pulses [15] allows tracking the complicated interplay of electronic interactions and lattice distortions that drive the phase transition. Furthermore, time-domain spectroscopy also offers a way to directly identify a symmetry broken state. This can be achieved by probing the state's collective excitations, namely the amplitude [16-21] and the phase mode [22, 23]. Probing the amplitude mode offers the possibility to map out the transient state's order parameter as a function of temperature and excitation density. For electronically driven transitions these measurements reveal a characteristic behavior of the electronic Higgs mode that allows distinguishing it from impulsively excited coherent phonons.

Here we study the collective excitations in $Ta_2NiSe_5$, a potential excitonic insulator, to prove the existence of a coherent condensate. In doing so we identified a low frequency phonon that under non-linear excitation directly couples to the excitonic condensate. This coupling arises from the quasi-1D structure of $Ta_2NiSe_5$, which consists of Ni and Ta chains aligned along the a-axis (figure 1a)). Recent ARPES measurements showed that the valance band flattening, characteristic of the excitonic insulator transition, occurs below 328K [7,8,24,25]. This is accompanied by a structural transition from an orthorhombic to a monoclinic crystal system, where the slight distortion does not result in the formation of a CDW [24,26]. Numerical calculations have shown that the band flattening and the structural transition can be attributed to the formation of excitons between Ta 5d electrons and hybridized Ni 3d-Se 4p holes [24]. An illustration of a charge transfer exciton in the material is shown in figure 1b): A completely filled valance band corresponds to a double occupancy of each Ni-site. An exciton is formed if an electron hops to the conduction band, supplied by the Ta-chains, and binds to the hole it left behind on the Ni-chain. This spatial separation of the electrons from the holes prevents recombination and is believed to be the reason for the stable exciton states in the material [26,27]. The consequence of the exciton formation and their BEC-like condensation is the opening of a large gap ($\Delta_{EI}$=160 meV) [25-27]. As the size of the gap directly denotes the frequency of the excitonic condensate's expected Higgs mode equivalent, it has a period ($\tau_\Delta$) of about $\tau_\Delta$= h/$\Delta$=26fs. In real-space this amplitude oscillation of the order parameter corresponds to a collective hopping of electrons and holes between the Ta and Ni chains (grey lines in figure 1b). However, it has been predicted recently that in the presence of strong electron phonon coupling the Higgs mode can also couple to a phonon, thus forming a new amplitude mode, which consist of intertwined lattice and order parameter oscillations [28]. Here we report strong evidence for the existence of such a coupled amplitude mode in Ta$_2$NiSe$_5$. This new collective mode is clearly distinguished from normal coherent phonons by its temperature and excitation density dependence. Its mode amplitude traces the behavior that would be expected for the purely electronic Higgs mode. This enables us to map out the order parameter as a function of temperature and excitation density.

In order to investigate the collective excitations of Ta$_2$NiSe$_5$ we performed pump-probe experiments using 1.55 eV (800 nm), 130 fs laser pulses. We probe the system's coherent oscillations below and above the critical temperature to characterize the temperature and excitation density dependence of their potential coupling to the excitonic condensate. To excite the modes without depleting the condensate significantly, we performed the experiments at perpendicular polarization to the chains. In the case of parallel polarization, the coupling of the light pulses to the condensate is enhanced. In this configuration a complete depletion of the condensate occurs even at low excitation densities (see supplemental material).

Figure 2a) depicts the time-dependence of the photoinduced reflectivity changes ΔR/R(t) at perpendicular polarization to the chains at different temperatures. The measured time-trace is a superposition of the electronic signal and several coherent oscillations. The electronic part describes the exciton breaking dynamics into electrons and holes and their subsequent recombination. It consists of a steep onset (time constants: ~0.1ps and ~1ps), and a double exponential decay (timescales ~1ps and ~10ps). The coherent oscillations on top of the electronic signal are made up of one mode at ~1 THz that we find to be coupled to the excitonic insulator (see discussion below) and two coherent phonons at higher frequencies, which were excited via impulsive stimulated Raman scattering [31]. To extract the coherent oscillations, we subtracted the fit to the electronic background (dotted black lines) from the measured signal. This yielded the oscillations shown in the inset of figure 2b). The corresponding

FFTs at 80K and 350K are presented in the main panel of figure 2b). It can be seen that at 80K the spectrum contains three distinct modes at 1,01 THz, 2,07THz and 2,98 THz (from here on denoted as 1 THz, 2 THz and 3 THz modes). Using LDA calculations and Raman measurements (see supplement) the 1THz oscillation was identified as an $A_{1g}$ mode, whereas the 2THz (3THz) mode is a $B_{1g}$ ($A_{1g}$) phonon. Comparing the FFTs at the different temperatures reveals that upon heating, a significant decrease in the peak height occurs for both the 1THz and 3THz mode, whereas the 2THz mode becomes heavily damped and disappears completely. To investigate the behavior of the modes quantitatively, we applied FFT band pass filters with a bandwidth of 1 THz around the respective peaks. The results for the 1THz mode, which are depicted in figure 2c), show a decrease of the amplitude with increasing temperature. To investigate this dependency systematically, we determined the initial oscillation amplitude $A_{EI}=\Delta R_{1THz}/R(t=0)$ by fitting a damped harmonic oscillator to the data.

Measuring $A_{EI}$ as a function of temperature and excitation density reveals that the 1 THz mode is indeed coupled to the excitonic condensate. The behavior of an uncoupled coherent phonon is presented, for comparison, on the example of the 3THz mode. Figure 3a depicts $A_{EI}$ as a function of temperature at different excitation densities. At the two highest excitation densities the amplitude remains almost constant at low temperatures. But on increasing the temperature the amplitude decreases with a behavior that is reminiscent of a typical mean field order parameter. The temperature dependence is completely different for normal coherent phonons in semiconductors, which show a behavior that is similar to the uncoupled phonon mode at 3 THz (figure 3b) or the equivalent non-coupled 1.2THz phonon in $Ta_2NiS_5$ (supplement). Figure 3b shows, on the example of the 3THz mode, that the amplitudes of these modes increase with increasing temperature. To show that the amplitude of the 1THz mode in Ta$_2$NiSe$_5$ (A$_{EI}$) indeed follows an order parameter behavior at a fixed excitation density, we fitted a BCS-type interpolation formula, which is also valid for the excitonic insulator [5], to the low temperature data: $\Delta(T) = \Delta(0) \tanh(\beta \sqrt{\frac{T_c}{T} - 1})$ [32], where $T_c$ denotes the critical temperature. Except for the excitation density dependent shift of the critical temperature to lower temperatures, this behavior is in agreement with the temperature dependence of the order parameter extracted from ARPES measurements and VCA calculations [25]. This directly underpins the coupling of the 1THz mode to the excitonic insulator state. At temperatures above T$_c$ the amplitude deviates from the order-parameter behavior and levels off into a regime, in which the amplitude only changes slowly but remains finite. This can be attributed to strong fluctuations, which are expected in the BEC-like regime of the excitonic insulator phase [6,8] and which were also observed in ARPES measurements [25]. In the supplementary material we show that at parallel polarization to the chains it is possible to completely deplete the condensate, which can be seen from a complete elimination of the 1THz mode (A$_{EI}$=0). At the excitation density of 0.35mJ/cm$^2$ (0.8mJ/cm$^2$) the fit ansatz yielded a T$_c$ of 302K (279K). The deviation of T$_c$ from the equilibrium value of 328K obtained from transport and ARPES measurements [25] can be attributed to the pump pulses, which break a certain number of excitons and hence deplete the excitonic insulator. Since the amount of optical depletion increases with increasing power, this also explains the shift of T$_c$ to lower temperatures with increasing excitation density. In contrast, in the low temperature regime, where thermal depletion doesn't play a role, stronger pump pulses lead to a greater excitation of the coupled exciton phonon system and therefore to a higher amplitude.

In order to determine the behavior of the coupling closer to equilibrium conditions, we investigated the temperature dependence also at a lower excitation density. The results, that are also depicted in Figure 3a, show that in this case the amplitude only follows the order parameter at temperatures above 225K. The corresponding fit yielded a $T_c$ of 330K, which, as expected for the low power case, agrees well with the equilibrium value of 328K. In the low temperature regime, however, the amplitude deviates from the order parameter behavior and instead of a constant value, an increase of the amplitude with increasing temperature is seen. The mode's linear Raman response (supplement) and the temperature dependence of the 3 THz coherent phonon (figure 3b) show that this is the behavior of an uncoupled coherent phonon. The fact that the mode behaves like an uncoupled phonon at low excitation densities indicates that there is an excitation threshold that has to be overcome in order to reach the non-linear excitation regime, in which the condensate couples to the phonon. This threshold is temperature dependent: At low temperatures the excitonic insulator is trapped in a deep potential and therefore a high excitation density is needed to overcome the threshold. However, at higher temperatures the thermal depletion of the excitonic condensate lowers the potential barrier and lower excitation densities are consequently sufficient to drive the excitonic system into the highly non-linear regime. In the supplementary material we show, that at even lower excitation densities the coupling remains inefficient over the entire temperature range and that the lack of coupling in these cases is also confirmed by the temperature dependence of the mode's frequency. In the coupled case, however, the amplitude $A_{EI}$ follows the order parameter of the excitonic insulator.

To further elaborate on the threshold behavior and characterize the excitation density dependence of the order parameter, we carried out power dependent measurements at two different temperatures: (1) at 120K, i.e. deep inside the excitonic insulator phase, and (2) at 250K, i.e. in a regime where thermal excitation of the condensate sets in (figure 4a). Both measurements show qualitatively the same behavior: At low excitation densities (shaded area) a steep onset is observed. In this regime the amplitude is proportional to the square root of the pump power, i.e. proportional to the electric field. With increasing power the amplitude deviates from this simple behavior, reaches a maximum and decreases roughly linearly when the excitation density is further increased. As discussed below this power dependence is a direct consequence of the coupling of the 1 THz phonon to the excitonic condensate. The 3THz mode's amplitude (see figure 4c), in contrast, increases linearly with power over the entire range, as expected for a coherent phonon [31-34]. The fits to the 1 THz mode's amplitude in figure 4 were accordingly constructed as a sum of the two components: a square root increase (component $A_1$) and a linear decrease with power (component $A_2$):

$$A_{EI}(\rho) = \underbrace{a \times \sqrt{\rho - \rho_c}}_{A_1} - \underbrace{b \times \mathrm{erf}\left(\frac{\rho - \rho_s}{\Delta\rho}\right)(\rho - \rho_c)}_{A_2},$$

where $\rho$ denotes the excitation density, and $\rho_s$ ($\Delta\rho$) characterizes the position of the threshold (the width). The factors a,b furthermore describe the amplitudes of the two components, and $\rho_c$ denotes the onset of the oscillation. The linear decrease in the fit ansatz was weighted by a step function to reproduce the relatively sharp transition between the two regimes. The necessity of including the step function, to properly match the experimental data, directly corresponds to the non-linear excitation threshold. This is illustrated in figure 4b, where the two components of the fit are plotted separately. The behavior of $A_2$ clearly shows that the linear decrease only sets in above a threshold ($\rho_s$). This can be understood as follows: At low powers the 1 THz phonon is not coupled to the condensate. But upon increasing the power the coupling sets in and the amplitude follows the transient order parameter. In this regime the decreasing amplitude indicates the increasing photo depletion of the condensate. The depletion increases linearly with the number of photons in the pump pulse, i.e. it depends linearly on the excitation density. The comparison of the two measurements at different temperatures shows that the threshold shifts to lower powers with increasing temperature, as emphasized by the two vertical lines in figure 4a. This also explains the previously discussed temperature dependence at the low excitation density (figure 3a). At low temperatures the excitation density is below the threshold. This prevents an efficient excitation of the coupled amplitude mode. But on increasing the temperature the threshold shifts to lower values, until it equals the excitation density. At this point the coupling becomes effective and the amplitude of the coupled exciton-phonon system ($A_{EI}$) follows the order parameter.

From the temperature and excitation density dependent measurements we can conclude that the 1THz oscillation indeed results from a coupling of the 1THz phonon to the excitonic insulator. Specifically, it can be described as a coupling to the Higgs mode of the excitonic condensate. This coupling can possibly be understood in a simplified picture: Our LDA calculations show that the 1THz phonon primarily modulates the Ni-Se distance (see supplement). As a result the Ni 3d - Se 4p hybridization is modulated [24], which in turn changes the exciton onsite energies and the spatial extension of the charge transfer excitons. This modulation most likely induces the collective hopping between the chains that characterizes the Higgs mode. Figure 5 visualizes the excitation mechanism of this coupled Higgs-phonon mode. At negative time delays (Fig. 5a) no prominent coupling between the EI and phonon potential is found. However, the pump pulse induces a coupling between EI and phonon (Fig. 5b): It depletes the exciton condensate and thus changes the excitonic insulator's potential on a time scale of 130fs, which is much longer than the 26fs period of the Higgs mode. Since the Higgs mode's period denotes the order parameter's intrinsic response time, the order parameter can follow the changing potential adiabatically, in the sense that the order parameter always remains in the minimum of the potential. The coupled phonon, however, has a much slower response time (1ps) and therefore can't follow this change adiabatically. This leads to an impulsive excitation (Fig. 5c) of the coupled Higgs-phonon system.

Investigating this coupled Higgs-phonon mode allows tracing the excitonic insulator's transient order parameter at the easily accessible frequency of the phonon. The measurements at low excitation densities show that this Higgs-phonon coupling only sets in above a certain threshold. This rules out a simple phonon excitation. Additionally, the 1 THz mode's decreasing amplitude at high excitation densities is a clear indicator of a photo-depletion of the excitonic condensate. We note that the excitation density dependence of the coupled mode's amplitude is independent of the electronic background signal, which shows a linear excitation density dependence running into saturation and no depletion (see supplement). This rules out a simple coupling of an impulsively excited coherent phonon to the electronic signal. The 1 THz mode's amplitude clearly follows the characteristic behavior of the Higgs mode. Its amplitude is proportional to the electric field of the pump pulses and decreases linearly with power at high excitation densities due to the photo-depletion. This behavior points directly to the excitation mechanism of the coupled Higgs-phonon system described above.

The experimental discovery of this mode in the nonlinear excitation regime of $Ta_2NiSe_5$ provides a first direct fingerprint of a coherent excitonic insulator ground state in the system. The collective mode's amplitude follows the system's transient order parameter as function of excitation density and temperature. The fact that probing the Higgs-phonon mode can identify a condensate is also confirmed by the behavior of the equivalent mode in $Ta_2NiS_5$, a material, which also shows strong excitonic binding [27], but no coherent condensate. In this material the 1THz mode shows the temperature and excitation density dependence of an uncoupled phonon (see supplement). This sensitivity to the coherent condensate makes the Higgs-phonon mode an ideal probe to investigate the suppression of the condensate upon S-doping in $Ta_2Ni(Se_{1-x}S_x)_5$ or the BEC-BCS-like crossover within the excitonic insulator phase that can be controlled in $Ta_2NiSe_5$ by tuning the band gap under pressure [26]. Beyond that, the novel scheme of probing a Higgs-phonon coupled mode opens the possibility to directly investigate electronically driven phase transitions that are not accompanied by charge density waves. It allows probing the coherent amplitude response at low frequencies. This can prove useful in systems with large electronic gaps, where a direct excitation of the coherent electronic system would be very challenging due to the required extremely short time scales.

**Acknowledgements:**

The authors thank A. Schulz for measuring the linear Raman spectra. We thank T. Larkin and A.V. Boris for sharing insights from optical measurements and fruitful discussions. We also thank T. Oka, S.R. Clark and M. Knap for helpful comments on the topic. We acknowledge support by the BW Stiftung and the MWK Baden-Württemberg through the Juniorprofessuren-Programm.

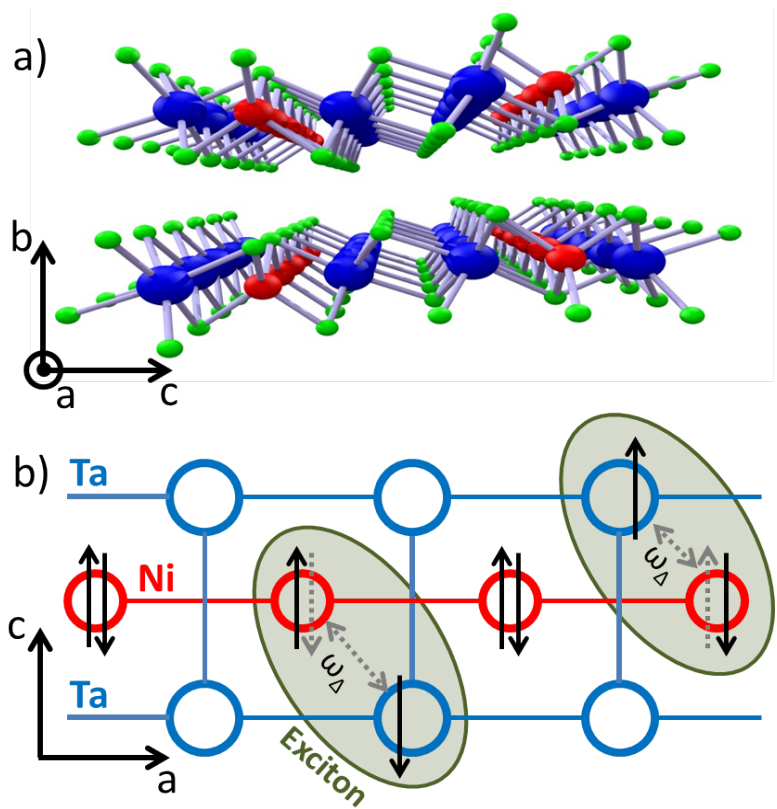

Figure 1: (a) Structure of Ta$_2$NiSe$_5$: The Ni and Ta 1D chains are aligned along the a-axis and forming sheets in the ac-plance. The electronic transport form along the chains in a-direction. (b) Structure and exciton formation along the chains: The Ni chains supply the valence band and the Ta chains the conduction band. In the semiconducting phase all Ni sites are doubly occupied. An exciton is formed between an electron on the Ta chains and a hole on the Ni-chain. The Higgs mode (frequency $\omega_\Delta$) corresponds to a collective hopping of electrons and holes between the chains.

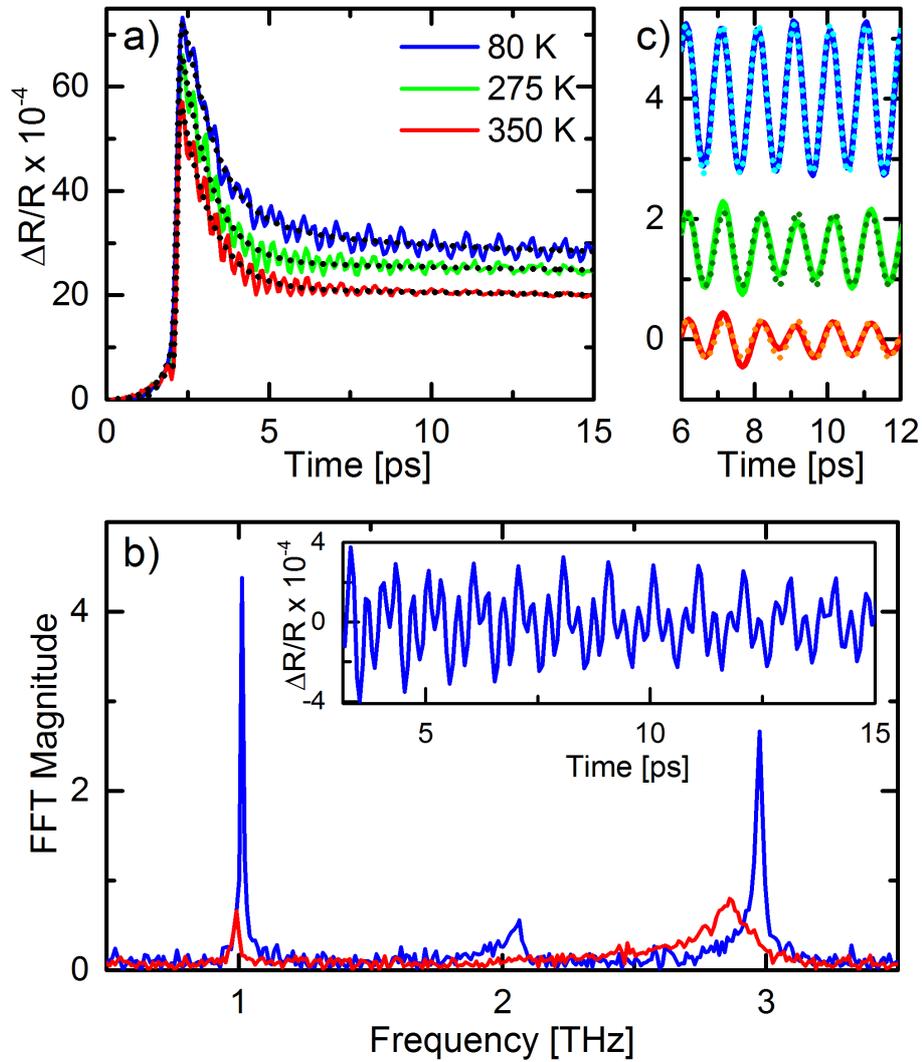

Figure 2: (a) Time-trace of photoinduced reflectivity changes at different temperatures and an excitation density of 0.35mJ/cm². The signal is made up of the electronic response and the coherent oscillations. The dotted black lines represent fits to the measured data. The inset of (b) shows only the coherent oscillations, which were extracted by substracting the fits in (a) from the measured data. The main panel of (b) presents the corresponding FFTs at 80K and 350K. (c) depicts the Higgs-phonon mode at 1 THz, which was extracted using a FFT band pass filter. The amplitude ($A_{El}$) of the coupled mode was determined by fitting a damped harmonic oscillator to the data (dotted lines).

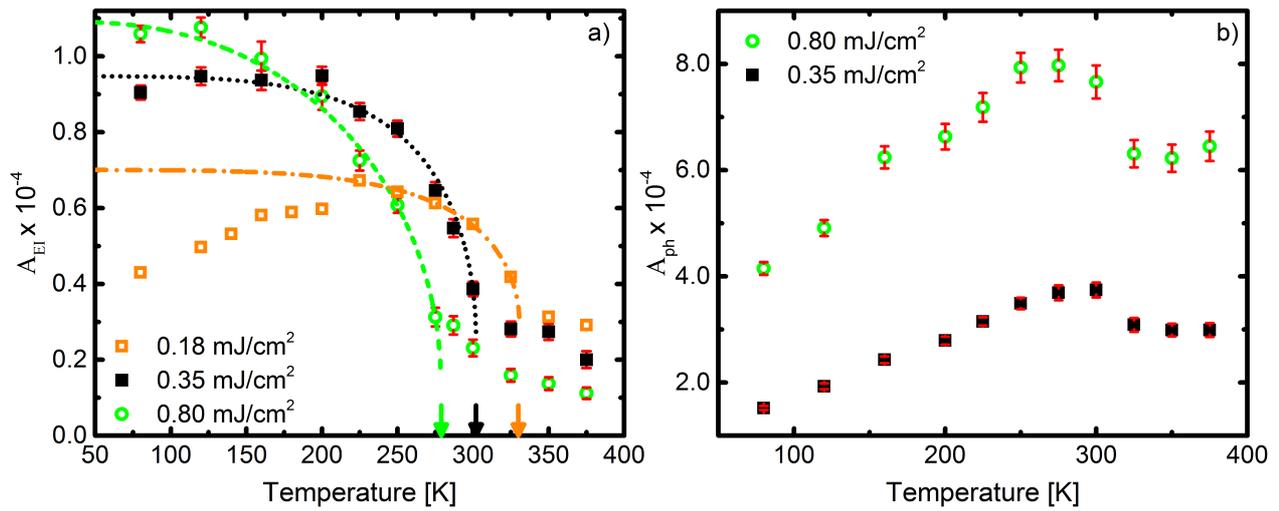

Figure 3: (a) Amplitude of the coupled mode ($A_{EI}$) at 1THz over temperature at different excitation densities. The fits (dotted and dashed lines) denote a mean field-like order parameter that was fitted to the low temperature data points. For the measurement at 0.18mJ/cm$^2$ only points above 225K were used for the fit. The arrows illustrate the position of $T_C$ at the respective excitation density. (b) Amplitude of the uncoupled phonon mode ($A_{ph}$) at 3THz over temperature at different excitation densities.

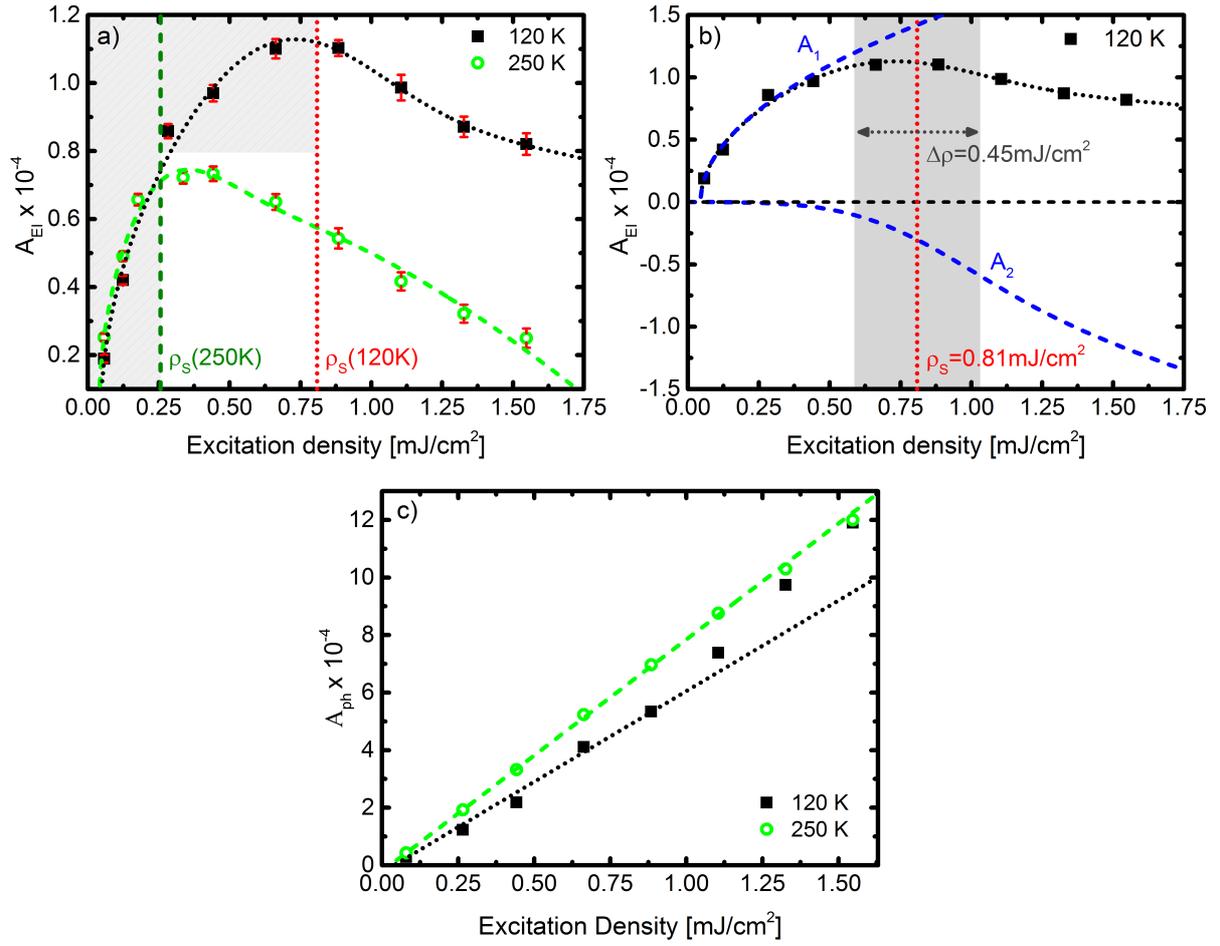

Figure 4: (a) Amplitude of the coupled mode ($A_{EI}$) at 1 THz over excitation density at 120K and 250K. As discussed in the text the fit (dotted and dashed lines) reveals the threshold ($\rho_s$) that characterizes the onset of the coupling to the excitonic condensate. The shaded area indicates the regime in which the coupling of the Higgs-mode to the phonon is not effective. (b) Separate fit components ($A_1$ and $A_2$) for the measurement at 120K. The shaded area describes the width of the step function, which characterizes the threshold. (c) Amplitude of the uncoupled phonon mode ($A_{ph}$) at 3 THz over excitation density at 120K and 250K.

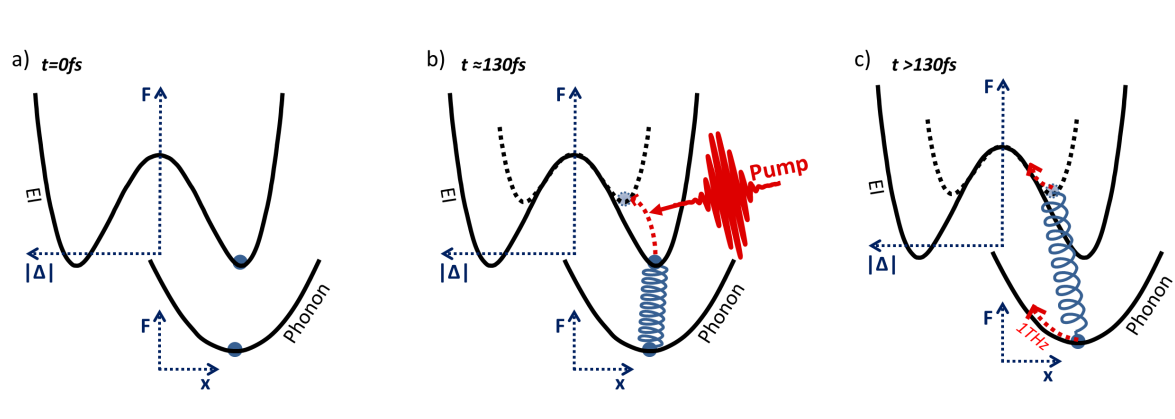

**Figure 5: Coupled EI and phonon potentials:** The double well potential represents the Excitonic Insulator (EI), and the single well potential the phonon. Due to the strong electron phonon coupling, a new amplitude mode emerges, which combines phonon and order-parameter dynamics. The excitation mechanism can be understood in the following: a) At negative time delays no coupling between the potentials can be observed. b) When the pump pulse arrives it changes the EI's potential energy landscape adiabatically, i.e. without exciting the Higgs mode directly. The potential shrinks and the order parameter reduces and a coupling (represented as spring) between the EI and the phonon becomes visible. c) This leads to an impulsive excitation of the 1THz phonon, since the change occurs faster than its intrinsic 1ps response time. Through the coupling to the excitonic insulator this results in an oscillation of the coupled Higgs-phonon system.

**Supplementary material to:**

**Coherent Order Parameter Oscillations in the Ground State of the Excitonic Insulator $Ta_2NiSe_5$**

Daniel Werdehausen[1,2], Tomohiro Takayama[1,3], Marc Höppner[1],

Gelon Albrecht[1,2], Andreas W. Rost[1,3], Yangfan Lu[4], Dirk Manske[1],

Hidenori Takagi[1,3,4], and Stefan Kaiser[1,2,*]

[1]*Max-Planck-Institute for Solid State Research, 70569 Stuttgart, Germany*

[2]*4th Physics Institute, University of Stuttgart, 70569 Stuttgart, Germany*

[3]*Institute for Functional Matter and Quantum Technologies,*

*University of Stuttgart, 70569 Stuttgart, Germany*

[4] *Department of Physics, The University of Tokyo, Bunkyo-ku,*

*Tokyo 113-0033, Japan*

*Corresponding author. Email: s.kaiser@fkf.mpg.de

## S1. Experimental and theoretical methods

**Sample preparation.** Single crystals of $Ta_2NiSe_5$ were grown by chemical vapour transport reaction. Elemental powders of tantalum, nickel and selenium were mixed with a stoichiometric ratio and sealed into an evacuated quartz tube ($\sim 1\times 10^{-3}$ Pa) with small amount of $I_2$ as transport agent. The mixture was sintered in a two-zone furnace under a temperature gradient of 900/850 ºC for a week. Thin strip-shaped crystals spreading in the *ac*-plane were grown at the cold end of tube. The crystals were characterized by x-ray diffraction and resistivity measurements.

**Pump-probe experiments.** We performed the time resolved experiments using a regenerative Ti:Sa amplifier delivering 6 μJ pulses at 800 nm with a pulse duration of 130 fs. The time resolved reflectivity changes were measured in a degenerate pump-probe experiment at this wavelength. The pump was attenuated to obtain the desired excitation density. The size of the pump spot was 300 μm and excitation densities up to 1.6 mJ/cm$^2$ were used. Higher excitation densities irreversibly damaged the samples. The samples were kept in an optical cryostat that allowed cooling to 80 K using liquid nitrogen and heating to 375 K.

**Raman measurements.** The linear Raman spectrum was measured using a Jobin Yvon Typ V 010 Labram single grating spectrometer, equipped with a double super razor edge filter and a peltier cooled CCD camera. The resolution of the spectrometer (grating 1800 L/mm) was 1 cm$^{-1}$. The spectra were taken in a quasi-backscattering geometry using the linearly polarized 632.817 nm line of a He/Ne gas laser. The power was lower than 1 mW and the spot size was 10 μm. The scattered signal was filtered analysed using an additional Polarizer before the spectrometer.

**Phonon spectrum.** To analyze the Raman spectrum and identify the oscillatory modes we calculated the phonon spectrum at q=(0,0,0) within the density functional perturbation theory[1] framework using the density functional code *quantum espresso*[2]. We used PAW pseudopotentials[3] from the PSLibrary v.0.3.1[4] together with plane waves as a basis set and employed the local density approximation in the parametrization of Perdew/Zunger[5]. The wave function (charge density) cutoff was set to 50~Ry (400~Ry) and a k-mesh of 4x1x4 was used. The forces (< 1~mRy/bohr) and the internal stress (< 0.5~kbar) of the initial crystal structure [6] were minimized prior to the lattice dynamics calculation.

## S2. Phonon Spectrum of Ta$_2$NiSe$_5$ and characterization of the 1 THz phonon

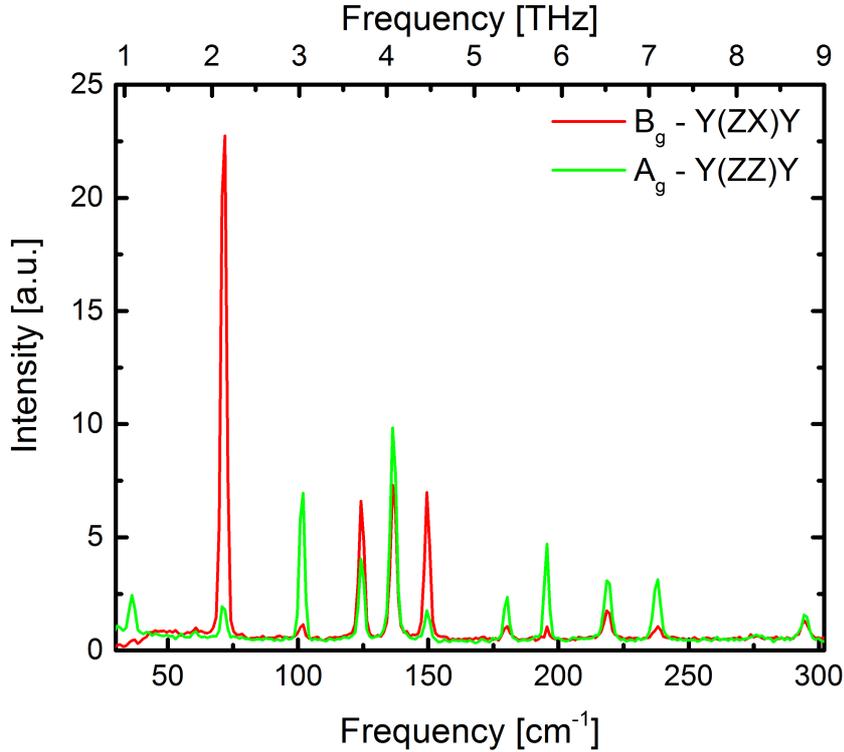

**Figure S2.1: Linear Raman measurements at different analyzer settings. The three lowest modes represent the ~1THz (36 cm$^{-1}$), ~2THz (72 cm$^{-1}$) and ~3THz (102 cm$^{-1}$) mode. The different analyzer settings show that the 1THz and 3THz mode have A$_{1g}$ symmetry, whereas the 2THz mode has B$_{1g}$ symmetry.**

Figure S2.1 presents the results of linear Raman measurements on Ta$_2$NiSe$_5$. A linearly polarized laser and an analyzer were used to identify the symmetry of the modes. The three lowest modes correspond to the 1THz, 2THz and 3THz modes, which were also observed in the time-resolved measurements. The two different analyzer settings reveal that the 1THz and 3THz modes have A$_{1g}$-symmetry, whereas the 2THz mode is a B$_{1g}$ phonon.

Table 1 lists the phonon spectrum to identify the modes. As described in the methods section the spectrum was obtained through LDA calculations. We note that the calculations assume the system to be a bad metal, which doesn't possess a pronounced gap. The 1THz mode can be clearly identified as the A$_{1g}$ mode at 0.94 THz (31.4 cm$^{-1}$). Figure S2.2 presents the corresponding mode's real-space atomic displacement. It can be seen that a block, consisting of a Ni-chain and the neighboring two Ta-chains, oscillates in a common direction with only a very small angle between the Ni and Ta displacements, whereas the neighboring block moves in the opposite direction. Therefore the relative distances between the Ni- and Ta-chains do not change significantly within the block. However, the Se atoms clearly change their position in respect to the Ni and Ta atoms. Due to the highly hybridized bands, this changes the band-structure and the onsite energies. This will affect the exciton binding energies and is therefore most likely the reason for the coupling of the phonon to the Higgs-mode.

|    | Frequency [THz] | Sym. | Act. |    | Frequency [THz] | Sym. | Act. |
|----|-----------------|------|------|----|-----------------|------|------|
| 1  | 0.94 | $A_g$    | R | 24 | 4.92 | $B_{1u}$ | I |
| 2  | 1.04 | $B_{1u}$ | I | 25 | 4.93 | $A_u$    | I |
| 3  | 1.24 | $B_{1g}$ | R | 26 | 5.30 | $A_g$    | R |
| 4  | 1.38 | $A_u$    | I | 27 | 5.30 | $B_{1u}$ | I |
| 5  | 1.49 | $A_g$    | R | 28 | 5.40 | $B_{1g}$ | R |
| 6  | 1.80 | $B_{1g}$ | R | 29 | 5.67 | $B_{1g}$ | R |
| 7  | 1.90 | $B_{1g}$ | R | 30 | 5.85 | $A_g$    | R |
| 8  | 2.27 | $B_{1g}$ | R | 31 | 6.14 | $A_u$    | I |
| 9  | 2.53 | $A_g$    | R | 32 | 6.25 | $A_u$    | I |
| 10 | 2.54 | $A_u$    | I | 33 | 6.35 | $B_{1u}$ | I |
| 11 | 2.57 | $A_g$    | R | 34 | 6.42 | $B_{1g}$ | R |
| 12 | 2.61 | $B_{1u}$ | I | 35 | 6.57 | $A_g$    | R |
| 13 | 2.90 | $B_{1g}$ | R | 36 | 6.65 | $B_{1u}$ | I |
| 14 | 2.97 | $B_{1u}$ | I | 37 | 6.72 | $B_{1g}$ | R |
| 15 | 3.26 | $A_u$    | I | 38 | 7.01 | $A_u$    | I |
| 16 | 3.27 | $B_{1u}$ | I | 39 | 7.02 | $B_{1u}$ | I |
| 17 | 3.32 | $B_{1g}$ | R | 40 | 7.14 | $A_g$    | R |
| 18 | 3.55 | $A_u$    | I | 41 | 7.14 | $B_{1g}$ | R |
| 19 | 3.75 | $A_g$    | R | 42 | 8.62 | $B_{1u}$ | I |
| 20 | 4.58 | $A_u$    | I | 43 | 8.68 | $B_{1g}$ | R |
| 21 | 4.59 | $B_{1u}$ | I | 44 | 9.06 | $A_u$    | I |
| 22 | 4.75 | $A_g$    | R | 45 | 9.07 | $A_g$    | R |
| 23 | 4.88 | $B_{1g}$ | R |    |      |          |   |

**Table 1: Phonon spectrum of Ta$_2$NiSe$_5$ in the low temperature phase at q=(0,0,0). The spectrum was obtained through LDA calculations. The column Sym. denotes the symmetry of the respective mode and the column Act. the activity (R: Raman; I: Infrared).**

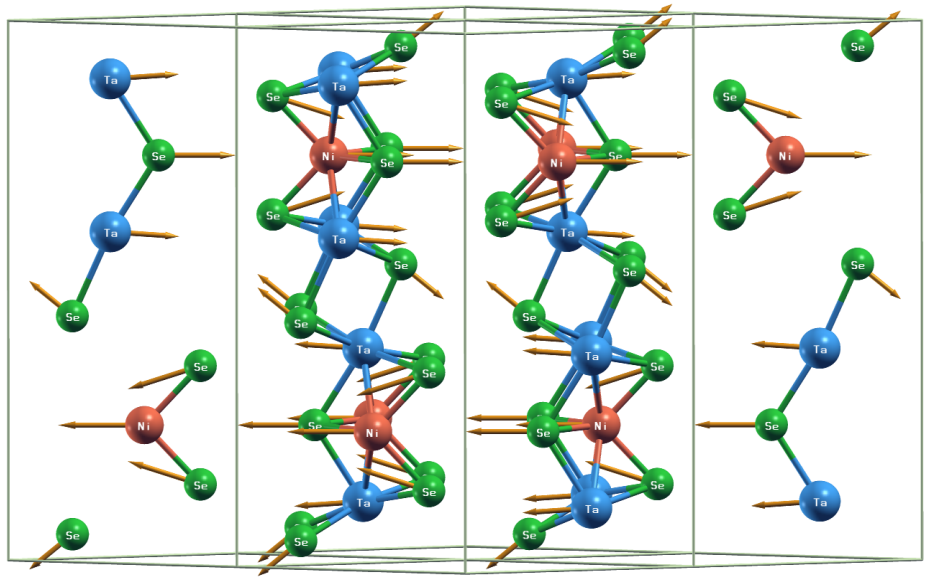

**Figure S2.2: Real space representation of the 1THz $A_{1g}$ phonon. One Ta-Ni-Ta block oscillates in a common direction, but two neighboring blocks move in opposite directions. Within a block the Se-Ni distance is modulated, which changes the highly hybridized bands.**

## S3.1 Amplitude behavior of the 1 THz mode at low excitation densities

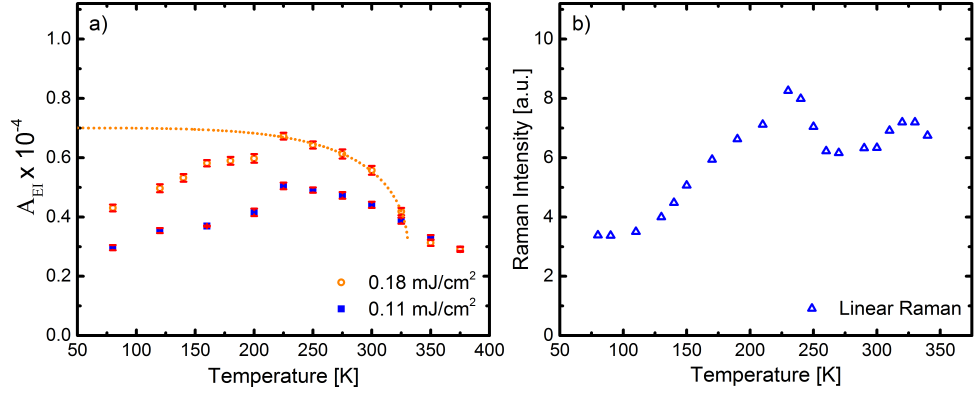

**Figure S3.1: Amplitude of the 1 THz mode at low excitation densities (a) and Raman Intensity (b) over temperature. The measurement at 0.18mJ/cm² is discussed in the main text. At 0.11mJ/cm² the excitation density remains below the threshold over the entire temperature range. The temperature dependence of the Raman intensity in b) qualitatively matches the behavior of the 3 THz mode (see S4), which shows that the mode behaves like a normal phonon under linear excitation.**

Figure S3.1a depicts the temperature dependence of the 1 THz mode's amplitude at low excitation densities. The behavior at 0.18mJ/cm² and the temperature dependent threshold for the coupling of the 1 THz phonon to the excitonic insulator are discussed in detail in the main text. The measurement at the lowest excitation density of 0.11mJ/cm² confirms that the coupling remains basically ineffective in this case. Although the behavior above 225K looks reminiscent of the order parameter, it doesn't show the characteristic drop close to the critical temperature. Consequently the mean-field like fit ansatz, which was fitted to high temperature data points at 0.18mJ/cm², fails to match the data at the low excitation density. This suggests that this excitation density remains below the threshold over the entire temperature range.

To investigate the mode's behavior at equilibrium we also carried out temperature dependent linear Raman measurements. Figure S3.1b presents the 1THz mode's scattered intensity over temperature. Its qualitative behavior matches the temperature dependence of the 3 THz mode's amplitude, which can be regarded as prototypical of a coherent phonon (see S4). This proofs that at equilibrium the mode is not coupled to the condensate.

## S3.2 Frequency shift of the 1 THz mode

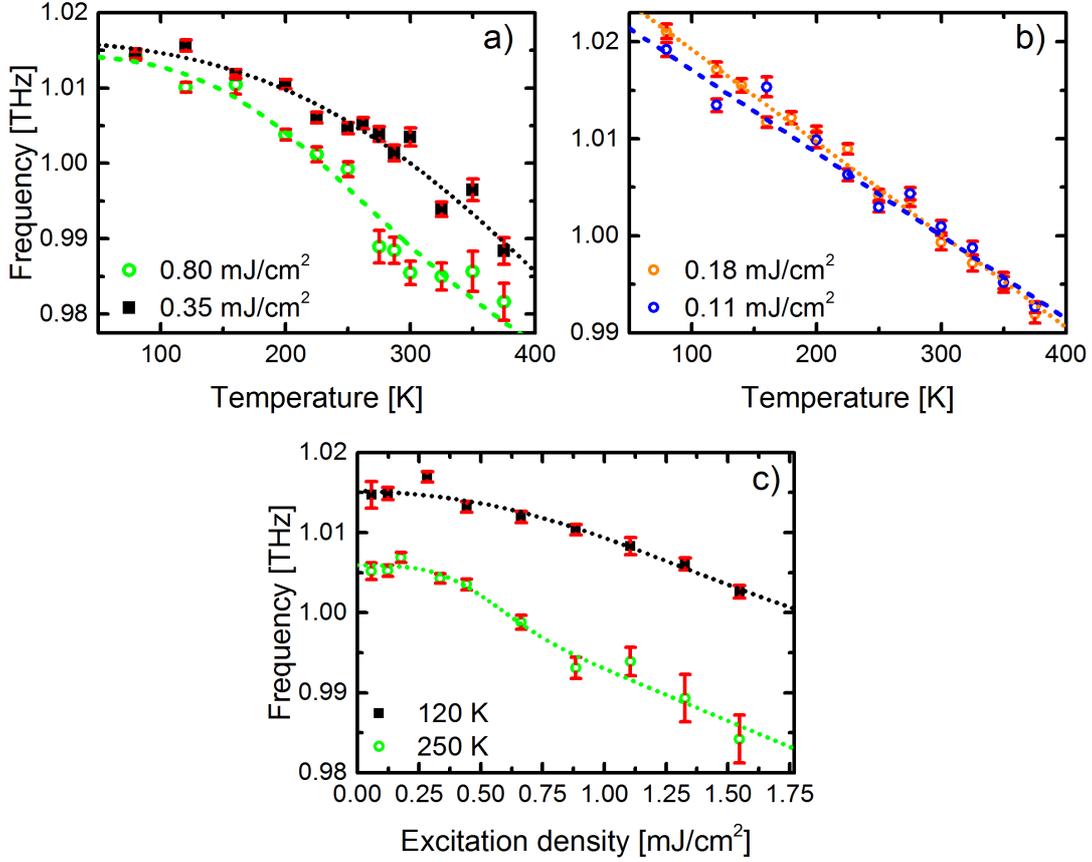

**Figure S3.2: Frequency of the coupled Higgs-phonon mode for all measurements discussed in the main text. (a) and (b) present the temperature dependence at the high and at the low excitation densities, respectively. At the low excitation densities the frequency exhibits a linear shift over the entire temperature range. This matches the behavior of the coherent phonon and therefore confirms the lack of coupling in this case. The power dependence is shown in (c).**

Figure S3.2 presents the 1 THz mode's frequency shift for all measurements discussed in the main text. The temperature dependence at the high excitation densities (figure S3.2a) shows, that in this regime the frequency decreases linearly at high temperatures and saturates towards low temperatures. However, at low excitation densities, where we do not see efficient coupling of the Higgs mode to the phonon, the behavior is linear across the entire temperature range. This supports the hypothesis that the coupling to the condensate is only effective at reasonably high densities. At low excitation densities the frequency shift is dominated by the phonon and therefore matches the behavior of a simple uncoupled coherent phonon like the 3 THz mode (see below - figure S4b). At higher densities the coupling to the condensate sets in, causing the frequency to saturate towards low temperatures. Figure S3.2c, shows that the saturation behavior occurs in the excitation density dependence. This behavior is again clearly distinguished from the 3 THz mode's linear power dependence, which is shown in figure S4c.

## S4. Coherent phonon - the 3 THz mode

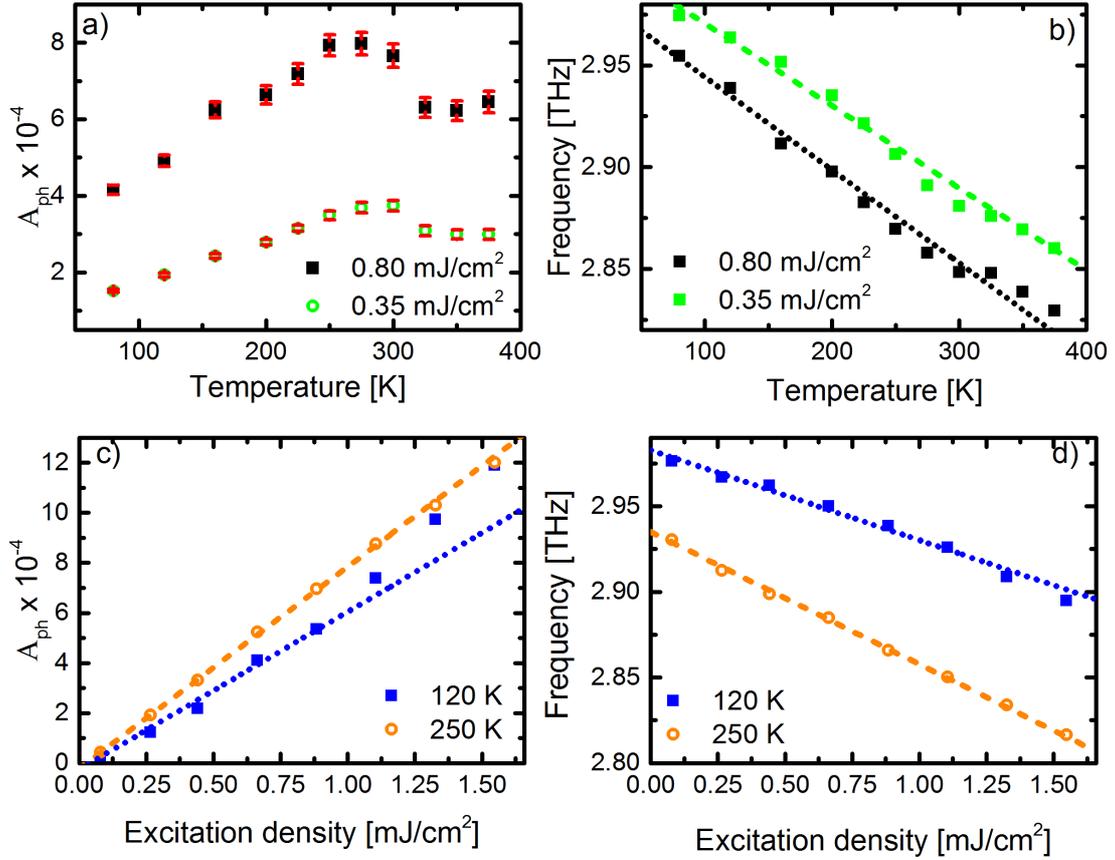

**Figure S4: Temperature (a) and excitation density dependence (c) of the 3THz coherent phonon's amplitude ($A_{ph}$). (b) and (d) present the corresponding temperature and excitation density dependence of the phonon frequency.**

Figure S4 presents the temperature and excitation density dependence of the 3THz mode's amplitude ($A_{ph}$) and frequency. Its behavior is clearly distinguished from the Higgs-phonon mode at 1THz, which is discussed in the main text and S3/S5. The results agree well with the behavior found for coherent phonons in other materials[7,8,9,10]. The temperature dependence in figure S4a shows that the amplitude increases linearly with increasing temperature up to a temperature of around 275K. At 300K a small drop occurs and the amplitude remains approximately constant when the temperature is further increased. This behavior is obviously not linked to the order parameter of the excitonic insulator. The temperature dependence of the frequency in figure S4b exhibits a roughly linear decrease, with a kink at 300K. The excitation density dependence depicted in figure S4c shows that the amplitude increases linearly with increasing pump power. The deviation from the linear behavior at 120K can be attributed to laser induced heating of the sample at high excitation densities.

---

[7] O. V. Misochko, et al., Phys. Rev. B **61**, 4305 (2000).
[8] B. Mansart et al., Phys. Rev. B **80**, 172504 (2009).
[9] K. Ishioka et al., J. Appl. Phys. **100**, 093501 (2006).
[10] K. Ishioka et al., Appl. Phys. Lett **89**, 231916 (2006).

## S5. Polarization parallel to the chains

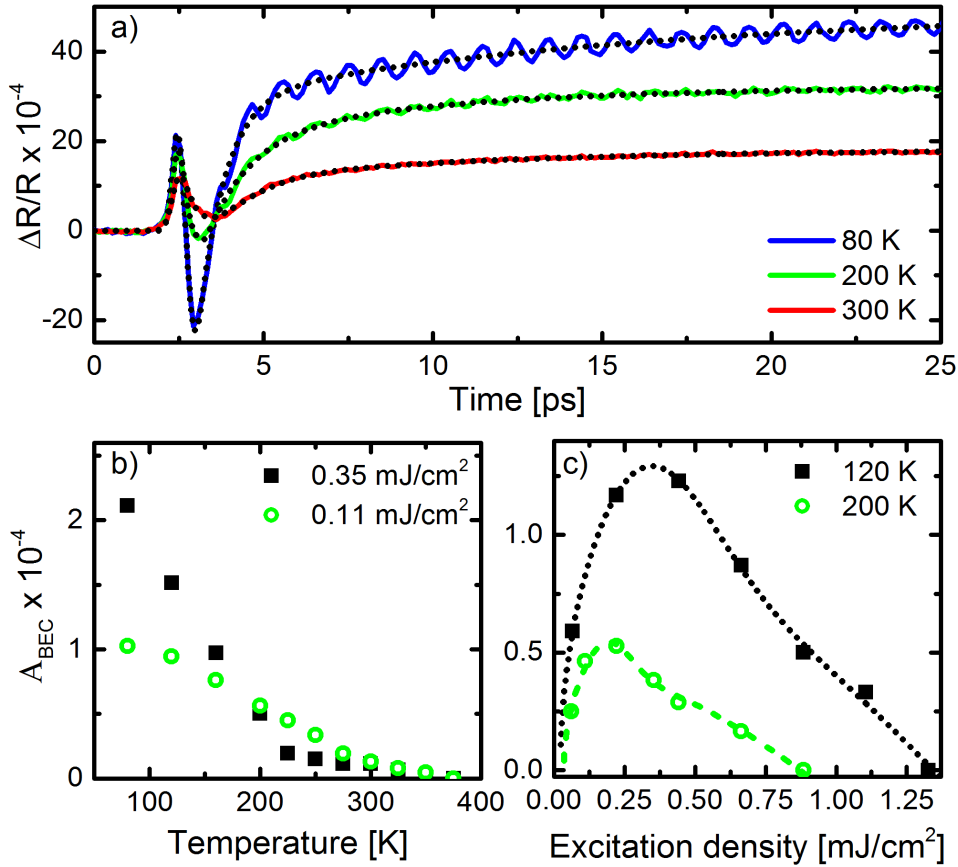

Figure S5: (a) Time-trace of photoinduced reflectivity changes at perpendicular polariazation to the chains. The dotted black lines represent fits to the electronic background of the measured data. (b) Temperature dependence and (c) excitation density dependence of the coupled mode's amplitude. In (c) the same fit ansatz as in the main part was used.

Figure S5a depicts the time-dependence of the photoinduced reflectivity changes at parallel polarization to the chains. At this polarization the transient signal shows a completely different behavior. The black dotted lines represent fits to the electronic background signal, which consist of an onset (time constant at 80K: 220fs) and three exponential decays (time constants at 80K: 50fs, 760fs, and 13ps). In this experimental configuration the oscillations on top of the signal contain only the 1 THz mode. We extracted the oscillations using the procedure discussed in the main text and performed the same systematic temperature and excitation density dependent measurements as described in the main text. Figure S5b presents the results for the temperature dependence of the 1 THz mode's amplitude ($A_{EI}$). It can be seen, that the qualitative behavior is identical to the case of orthogonal polarization. However, a more pronounced depletion of the condensate and a distinct shift of the drop to lower temperatures are observed. This suggests that the coupling of the pulses to the excitonic insulator is significantly enhanced at this polarization. This causes a significant depletion of the condensate even at low excitation densities. The excitation density dependence in figure S5c confirms this hypothesis. The qualitative behavior is again identical to the case of orthogonal polarization, but the enhanced coupling manifests itself in a pronounced shift of the peak and the depletion regime to lower excitation densities. At 120K the peak in the power dependence shifts by almost 90mW compared to the orthogonal polarization. At parallel polarization it was even possible to eliminate the amplitude mode completely ($A_{EI}=0$) by increasing the excitation density. This indicates a complete optical depletion of the excitonic condensate.

## S6. Coherent phonon oscillation of the 1 THz mode in $Ta_2NiS_5$

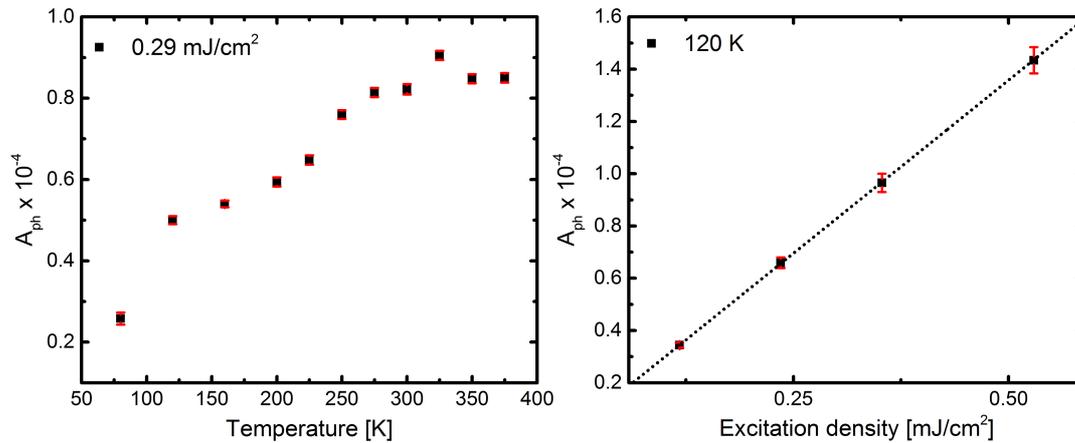

**Figure S6: Temperature (a) and excitation density dependence (c) of the 1THz coherent phonon amplitude ($A_{ph}$) in photoexcited $Ta_2NiS_5$.**

To confirm the sensitivity to the presence of an excitonic condensate we also investigated the related compound $Ta_2NiS_5$. This material also exhibits an optical gap with excitonic signatures[11] but does not possess an excitonic BEC[12]. In $Ta_2NiS_5$ the mode equivalent to the coupled 1 THz mode has a slightly higher frequency of 1.2THz. This can be attributed to the smaller mass of the sulfide atoms compared to the selenide atoms. Figure S6 presents the results of the corresponding measurements. As expected both the temperature dependence (a) and the excitation density dependence (b) reveal no signature of a condensate. They instead resemble the results for the uncoupled coherent phonon at 3THz (see S4). Higher excitation densities couldn't be used in this case, since the damage threshold of $Ta_2NiS_5$ is significantly lower than the one of $Ta_2NiSe_5$.

---

[11] T. Larkin, A. V. Boris, H. Takagi, private communication.
[12] Y. F. Lu, H. Takagi, private communication.

## S7. Excitation density dependence of the electronic amplitude

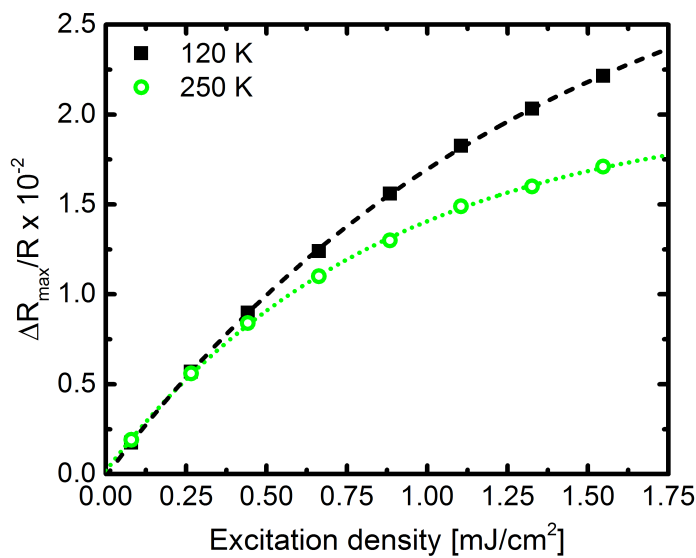

**Figure S7: Maximum of the electronic signal as a function of the excitation density at different temperatures.**

Figure S7 presents the excitation density dependence of the electronic signal's amplitude at different temperatures. The data points correspond to the maximum of the fit to the electronic signal. The figure shows that the amplitude increases linearly at small excitation densities and eventually approaches saturation at high excitation densities. This behavior is clearly distinguished from the excitation density dependence of the coupled mode's amplitude (see main text), which proves that the mode doesn't simply couple to the electronic background.